\documentclass[5p,a4paper]{elsarticle}
\usepackage{epsfig}
\usepackage{amsmath,amsfonts,amssymb}

\usepackage{xcolor}
\usepackage{colortbl}
\usepackage{pstricks,multido}

\definecolor{lightyellow}{rgb}{.95,.96,.5}
\definecolor{myblue}{rgb}{.61,.61,1}
\definecolor{midblue}{rgb}{.7,.7,1}
\definecolor{lightblue}{rgb}{.9,.9,1}
\definecolor{mygrey}{rgb}{.75,.75,.75}
\definecolor{lightred}{rgb}{1.,0.8,0.8}
\definecolor{lightgreen}{rgb}{0.66,1.0,0.67}

\renewcommand{\epsilon}{\varepsilon}
\def\la{\langle}
\def\ra{\rangle}
\def\be{\begin{equation}}
\def\ee{\end{equation}}
\newcommand\bea{\begin{eqnarray}}
\newcommand\eea{\end{eqnarray}}
\newcommand\equal{&\!\!\!\!=\!\!\!\!&}

\begin{document}

\title{Finite-size corrections for logarithmic representations in critical dense polymers}

\author[tai,yer,ntu]{Nickolay Sh. Izmailian}
\ead{izmailan@phys.sinica.edu.tw}

\author[lou]{Philippe Ruelle}
\ead{philippe.ruelle@uclouvain.be}

\author[tai]{Chin-Kun Hu}
\ead{huck@phys.sinica.edu.tw}

\address[tai]{Institute of Physics, Academia Sinica, Nankang,
Taipei 11529, Taiwan}
\address[yer]{Yerevan Physics Institute, Alikhanian Br. 2,
375036 Yerevan, Armenia}
\address[ntu]{National Center for Theoretical Sciences, Physics
Division, National Taiwan University, Taipei 10617, Taiwan}
\address[lou]{Institut de Recherche en Math\'ematique et Physique,
Universit\'e catholique de Louvain,
B-1348 Louvain-La-Neuve, Belgium}

\date{\today}

\begin{abstract}
We study (analytic) finite-size corrections in the dense
polymer model on the strip by perturbing the critical Hamiltonian
with irrelevant operators belonging to the tower of the identity.
We generalize the perturbation expansion to include Jordan cells,
and examine whether the finite-size corrections are sensitive to
the properties of indecomposable representations appearing in the
conformal spectrum, in particular their indecomposability parameters. 
We find, at first order, that the corrections do not depend on these parameters nor even on the presence of Jordan cells. Though the corrections themselves are not universal, the ratios are universal and correctly reproduced by the conformal perturbative approach, to first order.
\end{abstract}


\maketitle

\section{Introduction}
\label{introduction}

Universality, scaling, and finite-size corrections for critical
systems have attracted much attention in recent deca\-des
\cite{69jmpDimer,69prIsing,fishercorr,percoHu,percoZiff,izmailian2001,03preDimer,06preZiff}.
Although many theoretical results are now known about the critical
exponents and universal relations among the leading critical
amplitudes, not much information is available on ratios among the
amplitudes in finite-size correction terms \cite{fishercorr}. New
universal amplitude ratios have been recently presented for the
Ising model \cite{izmailian2001}. In this paper, we compute
finite-size corrections in the critical dense polymer model \cite{pearce2007}, and compare them with the results obtained in perturbation theory. Compared to previous analyses, the new feature here is
the presence of indecomposable representations in the conformal spectrum \cite{pearce2007,mdsa,vjs}, which first requires to adapt the perturbation
theory to include such representations. In particular it raises the question
as to whether the finite-size corrections depend on their indecomposability
parameters \cite{roh,gk,kr,pr,djs}.

Conformal invariance implies that on an infinitely long strip of
finite width $N$, the eigenstates of the critical transfer matrix,
associated to conformal states, have energies which scale with $N$
like \cite{Blote} 
\be 
E_n(N) = N f_b + f_{s} +
\frac{\pi\zeta(\Delta_n-c/24)}{N} + {\cal O}(N^{-2}), 
\label{En}
\ee 
where the bulk free energy density $f_b$, the surface free
energy $f_{s}$, and the anisotropy factor $\zeta$ are
non-universal constants; in contrast, the central charge $c$ and
the weights $\Delta_n$ of the conformal eigenstates, are universal
(although boundary dependent for $\Delta_n$).

Higher order correction terms in (\ref{En}) are nonuniversal;
however it has been suggested recently \cite{izmailian2001}, and
partially checked for the Ising universality class ($c=1/2$), that
the asymptotic expansion of the critical eigenvalues has the form
($n=0$ will refer to the groundstate) 
\be 
E_n(N) = N f_b + f_{s} + \sum_{p=1}^{\infty}\frac{a^{(n)}_p}{N^{2p-1}}, 
\label{Enexp} 
\ee
and that the amplitude ratios $a_p^{(n)}/a^{(0)}_p$ are universal
and depend only on the boundary conditions
\cite{izmailian2001,izmailian2009ab}. The case $p=1$ readily
follows from (\ref{En}).

For critical dense polymers, it is not difficult to see from the results
of \cite{pearce2007} that the finite-size energy levels have an asymptotic expansion 
like in (\ref{Enexp}). The general calculation of all coefficients
is a tedious exercise, which can however be completed for an infinite number
of levels (among these, the lowest-lying levels, as well as the lowest-lying
logarithmic pairs, in each representation). Moreover we check that the first
coefficients $a_2^{(n)}$ match the perturbative values, and that the
ratios $a_2^{(n)}/a^{(0)}_2$ are universal.

\section{Critical dense polymers}
\label{cdp}

We consider the model of critical dense polymers on a square grid, as formulated
in \cite{pearce2007}. The degrees of freedom are localized on the elementary faces 
of the lattice, which can be in two different configurations, see Eq. (2.1) of \cite{pearce2007}.
The model on a strip, of width $N$ and height $2M$, has a double-row transfer matrix ${\bf
D}(u)$. The partition function is given by \cite{pearce2007}
\be 
Z_{N,M} = {\rm Tr}\,{\bf D}(u)^M = \sum_n e^{-2M E_n(N;u)}, 
\ee 
where the sum is over all eigenvalues of ${\bf D}(u)$, written as
$e^{-2E_{n}(N;u)}$; $u$ is a spectral parameter related to the
anisotropy factor by $\zeta = \sin{2u}$.

The configuration space on which the transfer matrix acts can be
divided into sectors ${\cal L}_{N,\ell}$, labelled by an integer
$\ell \geq 0$, of the same parity as $N$. The integer $\ell$ is
the number of defect lines, and can be seen as
fixing a specific boundary condition. The range of $\ell$ is
finite when $N$ is finite, but will be considered as unbounded in
the limit $N \to \infty$. It was found in \cite{pearce2007} that
the transfer matrix is diagonalizable when restricted to each
sector separately, although analytic and numerical analyses
indicated that it is not in the full space when $N$ is even. This
was confirmed in \cite{mdsa,rasm}, where it was
shown that the sectors $\ell$ and $\ell+2$, for $\ell = 0 \bmod
4$, are linked up by rank 2 Jordan blocks, while there are no
Jordan cells when $N$ is odd.

All eigenvalues have been determined in \cite{pearce2007}, for any
finite value of $N$. They are given by 
\be 
E_n = -{\frac{1}{2}}\sum_{j=1 \atop j=N \bmod 2}^{N-2} \!\! \log\Big[(1 +
\zeta \,\epsilon_j \sin{ \frac{\pi j}{2N}})(1 + \zeta \, \mu_j
\sin{ \frac{\pi j}{2N}})\Big],
\label{en} 
\ee 
where the summation includes the integers $j$ with the same parity as $N$. The
eigenvalues also depend on parameters $\epsilon_j,\mu_j$ equal to
$\pm 1$, although not all choices of $\epsilon_j,\mu_j = \pm 1$
are allowed. Exactly which sequences of $+1,-1$ correspond to
actual eigenvalues, and for which sector, are given by the
selection rules conjectured in \cite{pearce2007}, and proved very recently in \cite{morin2011}. 
What we need is summarized below.

A main result from \cite{pearce2007} is that the set of
eigenvalues of the transfer matrix in the sector ${\cal
L}_{N,\ell}$ is such that it leads, through (\ref{En}), to a set
of conformal weights $\Delta_n$ whose values and degeneracies
exactly match those of a highest weight representation ${\cal
V}_{1,s}$ of weight $h_{1,s} = {(s-2)^2-1 \over 8}$, with $s=\ell+1$.
As a representation of the Virasoro algebra, ${\cal V}_{1,s}$ is
believed to be the quotient of the highest weight Verma module
$V_{1,s}$ by the singular vector at level $s$ \cite{rasm}.

The overall groundstate $E_0$, found in ${\cal L}_{N,0}$ or ${\cal
L}_{N,1}$, corresponds to all $\epsilon_j=\mu_j=1$ in (\ref{en}).
From the Euler-MacLaurin summation formula, the asymptotic
expansion of $E_0$ takes the form (\ref{Enexp}) with the
coefficients 
\be 
a_p^{(0)} =\frac {\pi^{2p-1} {\rm
B}_{2p}(\alpha)} {(2p)!} \, \lambda_{2p-1}(\zeta), \quad
\alpha=\frac{N}{2} \bmod 1,
\ee 
where B$_n(z)$ are the Bernoulli polynomials, $\alpha=0,{1 \over 2}$ and 
$\lambda_{2p-1}(\zeta)$ are polynomials of degree
$2p-1$, obtained from the following expansion,
\be
\log{(1 + \zeta \sin{x})} = \sum_{n=1}^\infty {\lambda_n(\zeta)\over n!} x^n.
\ee
The first few odd order polynomials read $\lambda_1(\zeta)=\zeta, \, \lambda_3(\zeta)
= 2\zeta^3-\zeta, \, \lambda_5(\zeta) = 24\zeta^5 - 20 \zeta^3 + \zeta$, see also 
\cite{nigro2009}.

For $p=1$ in particular, one finds $\Delta_0-c/24 = 1/12$ for $N$ even
and $-1/24$ for $N$ odd. Assuming $c=-2$, this gives $\Delta_0 = h_{1,1}
= 0$ for $N$ even, and $\Delta_0 = h_{1,2} = -1/8$ for $N$ odd.

For $p=2$ one finds from the previous formula \be a_2^{(0)} =
\frac{\pi^3 }{6} \big(\Delta_0^2 - \textstyle{ \frac{1}{120}}\big)
\,\lambda_3(\zeta). \label{ags} \ee

The excited levels are obtained by switching the $\epsilon_j,\mu_j$
from $+1$ to $-1$ in a way allowed by the selection rules \cite{pearce2007,morin2011}.
The specific levels we will consider here are the two lowest-lying levels
in each sector ${\cal L}_{N,\ell}$, denoted $E_0^{(\ell)}$ and $E_1^{(\ell)}$,
for all $\ell \geq 1$ except $\ell=2$ (which is somewhat special and requires
a separate treatment; the following checks however work for them too). They are
non-degenerate (within their sector) and are obtained by setting to $-1$ the
parameters $\mu_j$ with the following indices (keeping those of the appropriate parity),
\bea
&& 1 \leq j \leq \ell-2 \qquad \text{for }E_0^{(\ell)},\\
&& 1 \leq j \neq \ell-2 \leq \ell \qquad \text{for }E_1^{(\ell)}. 
\eea

Working out the asymptotic expansion of $E_r^{(\ell)}$ for
$r=0,1,$ yields the first coefficients (with no assumption on $c$)
\be 
a_1\big(E_r^{(\ell)}\big) = \pi \, \big(\textstyle h_{1,s} + r + \frac{1}{12}\big) \,
\lambda_1(\zeta), \label{a1}
\ee 
and 
\bea 
a_2\big(E_0^{(\ell)}\big) \equal \frac{\pi^3}{6} \big(h_{1,s}^2 - \textstyle{\frac{1}{120}}\big) \,
\lambda_3(\zeta), 
\label{a20}\\
a_2\big(E_1^{(\ell)}\big) \equal \frac{\pi^3}{6} \big(h_{1,s}^2 + 6\,h_{1,s} +
\textstyle{\frac{119}{120}}\big) \, \lambda_3(\zeta).
\label{a21} 
\eea 
Before comparing them with the perturbative
approach, it is useful to know where the Jordan blocks appear when
one considers the full configuration space, namely the union of
all sectors.

As mentioned above, Jordan cells appear only for $N$ even, in which case the
full configuration space is given by $\cup_{\ell \geq 0, {\rm even}}\, {\cal L}_{N,\ell}$.
It turns out that within each sector, the eigenstates are generically non-degenerate,
implying that the transfer matrix is diagonalizable in each sector separately;
however every sector ${\cal L}_{N,\ell}$ has a subset of eigenstates which are
pairwise degenerate with eigenstates of ${\cal L}_{N,\ell-2}$, while the
complementary subset is pairwise degenerate with eigenstates of ${\cal L}_{N,\ell+2}$;
these double degeneracies occur {\it for all values of $N$ (even) and $\zeta$}.

In the conformal limit (the order $N^{-1}$ in the expansion of the eigenvalues),
the degeneracy of the eigenstates increases. For $N$ odd, each sector $\ell$
odd carries an irreducible highest weight representation ${\cal V}_{1,\ell+1}$.
For $N$ even, it is believed \cite{pearce2007,mdsa,rasm} that the pairs of sectors
$\ell$ and $\ell+2$, for $\ell=0 \bmod 4$, and their associated representations
${\cal V}_{1,\ell+1}$ and ${\cal V}_{1,\ell+3}$, are tied together by rank 2
Jordan cells to form staggered modules \cite{roh,gk,kr}.

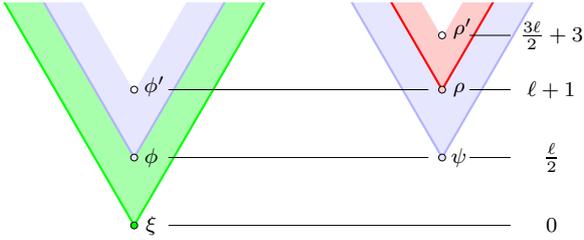
\begin{figure}[tbp]
\begin{center}
\begin{pspicture}[shift=0](0.5,2.2)(6.5,5.3)
\psclip{\psline[linestyle=none](-0.5,-0.)(-0.5,5.2)(7.5,5.2)(7.5,-0.)}
\psset{xunit=.9cm}
\psset{yunit=.9cm}
\psset{runit=.9cm}
\pswedge[linecolor=green,fillstyle=solid,fillcolor=lightgreen](1.5,2.5){5}{60}{120}
\pswedge[linecolor=midblue,fillstyle=solid,fillcolor=lightblue](1.5,3.5){5}{60}{120}
\pswedge[linecolor=white,fillstyle=solid,fillcolor=white](1.5,4.5){5}{60}{120}
\pscircle[fillstyle=solid,linewidth=0pt,fillcolor=green](1.5,2.5){0.05}
\pscircle[fillstyle=solid,linewidth=0pt,fillcolor=lightblue](1.5,3.5){0.05}
\pscircle[fillstyle=solid,linewidth=0pt,fillcolor=white](1.5,4.5){0.05}
\rput(1.75,2.5){\footnotesize $\xi$}
\rput(1.75,3.5){\footnotesize $\phi$}
\rput(1.8,4.55){\footnotesize $\phi'$}
\pswedge[linecolor=midblue,fillstyle=solid,fillcolor=lightblue](6,3.5){5}{60}{120}
\pswedge[linecolor=red,fillstyle=solid,fillcolor=lightred](6,4.5){5}{60}{120}
\pswedge[linecolor=white,fillstyle=solid,fillcolor=white](6,5.3){5}{60}{120}
\pscircle[fillstyle=solid,linewidth=0pt,fillcolor=lightblue](6,3.5){0.05}
\pscircle[fillstyle=solid,linewidth=0pt,fillcolor=lightred](6,4.5){0.05}
\pscircle[fillstyle=solid,linewidth=0pt,fillcolor=white](6,5.3){0.05}
\rput(6.25,3.5){\footnotesize $\psi$}
\rput(6.25,4.5){\footnotesize $\rho$}
\rput(6.3,5.4){\footnotesize $\rho'$}
\psline[linecolor=black,linewidth=0.1pt](2.0,2.5)(7.0,2.5)
\psline[linecolor=black,linewidth=0.1pt](2.0,3.5)(5.8,3.5)
\psline[linecolor=black,linewidth=0.1pt](6.4,3.5)(7.0,3.5)
\psline[linecolor=black,linewidth=0.1pt](2.0,4.5)(5.8,4.5)
\psline[linecolor=black,linewidth=0.1pt](6.4,4.5)(7.0,4.5)
\psline[linecolor=black,linewidth=0.1pt](6.4,5.3)(7.0,5.3)
\rput(7.6,2.5){\footnotesize $0$}
\rput(7.6,3.5){\footnotesize $\ell \over 2$} 
\rput(7.6,4.5){\footnotesize $\ell+1$}
\rput(7.6,5.3){\footnotesize ${3\ell \over 2}+3$}
\endpsclip
\end{pspicture}
\end{center}
\caption{(Color online) The cone-like figures represent the
representations ${\cal V}_{1,\ell+1}$ (left) and ${\cal V}_{1,\ell+3}$ (right).
The numbers on the right show the levels at which the various fields occur.}
\label{fig1}
\end{figure}

Fig. 1 schematically represents the two representations 
${\cal V}_{1,\ell+1}$ and ${\cal V}_{1,\ell+3}$. The left one has
highest weight state $\xi$, a first singular vector $\phi$, and a
second singular vector $\phi'$, set to zero, and similarly on the
right for $\psi,\,\rho$ and $\rho'$, the latter being also set to zero. The states
in the two blue regions are conformal states coming from
eigenstates of the transfer matrix which are pairwise degenerate
for all (large enough) values of $N$; they form logarithmic pairs
and are related by the non-diagonalizable action of $L_0$, like
f.i. $(L_0-\Delta)\phi = 0,\, (L_0-\Delta)\psi=\phi$.

The full projective representation, denoted by ${\cal R}_{1+\ell/2}$, spans the
two sectors ${\cal L}_{N,\ell}$ and ${\cal L}_{N,\ell+2}$, and can be partitioned into two
disjoint subsets : $S_1$ contains eigenstates of $L_0$ which are not parts of a logarithmic
pair (green and red regions), and $S_2$ contains logarithmic pairs (in blue).
Each such representation ${\cal R}_{1+\ell/2}$ comes with an intrinsic parameter
$\beta_{1+\ell/2}$, which fixes its equivalence class \cite{roh,gk,kr,pr}.

\section{Conformal perturbation theory}
\label{conformal}

\vskip 2 mm


The finite-size
corrections in (\ref{Enexp}) can in principle be computed in
perturbation theory \cite{cardy86}. In general, the critical
lattice Hamiltonian will contain correction terms to the
fixed-point Hamiltonian $H_c = \frac{\pi}{N}(L_0 - \frac{c}{24})$,
\begin{equation}
H = \zeta H_c + \sum_k \: g_k \int_0^N \varphi_k(v) \, {\rm d} v, \label{Hc}
\end{equation}
where the $\varphi_k$ are irrelevant quasi-primary fields, and $g_k$ are non-universal
constants. On general grounds, a perturbation $\varphi_k$ with scaling dimension
$x_k$ brings a first-order correction to the energy proportional to $N^{1-x_k}$.

Fields in the tower of the identity are available in any conformal theory.
On the strip, the lowest contributing field\footnote{The correct quasi-primary combination 
of dimension 4 is $L_{-2}^2(v) - \frac{3}{5} L_{-4}(v)$; however the terms $L_{-n}(v)$ for 
$n \geq 3$ give no contribution to any order in perturbation \cite{reinicke87}.} 
is $\varphi = L_{-2}^2(v)$, of dimension 4 \cite{reinicke87}; it corrects the conformal 
energies by a $N^{-3}$ term, and should reproduce the first correction coefficient
$a_{2}^{(n)}$ in (\ref{Enexp}).

Perturbation techniques have been studied for a long time, and successfully
applied in concrete models \cite{henkel}. Wri\-te $H = H_0 + gV$. A basic
result states that the first-order correction to the energy $E_{c,n}$ of
a non-degenerate eigenstate $|n\ra$ of $H_0$ is given by $\la n|gV|n\ra$.
Here it is sufficient that the state $|n\ra$ be non-degenerate within its
own representation since $\varphi$ preserves each representation.

This scheme can be readily applied to the critical dense polymer model when $N$ is odd,
or to the two lowest-lying states in the indecomposable representations
${\cal R}_k$ ($\xi$ and $L_{-1}\xi$) when $N$ is even, but should be
adapted for states in Jordan cells. In addition the representations
${\cal R}_k$ carry an intrinsic parameter $\beta_k$, which may affect the corrections.

It has been appropriately emphasized in \cite{djs,vjs} that the conformal
states should be equipped with a scalar product for which the conformal
Hamiltonian is hermitian. When the Hamiltonian is not diagonalizable,
this bilinear form cannot be positive definite and differs from the
canonical scalar product usually used in the lattice configuration space.
For a typical rank 2 Jordan cell in the subset $S_2 \subset {\cal R}_k$,
one can take the scalar product to be such that
$\la \phi|\phi\ra = 0, \,\la \phi|\psi\ra = \la \psi|\phi\ra = \beta$,
$\la \psi|\psi\ra = \gamma$ where $\beta=\beta_k$ if $(\phi,\psi)$ is
the lowest logarithmic pair of ${\cal R}_k$, otherwise $\beta$ is a
calculable multiple of $\beta_k$ for a descendant logarithmic pair \cite{vjs}.
For states in $S_1$, the usual scalar product is used.

Assume that $\phi,\psi$ are the conformal states of a non-degenerate
logarithmic pair (only pair at their level),
\be
(H_0 - \lambda_0) \phi = 0, \qquad (H_0 - \lambda_0) \psi = \phi,
\ee
with $\lambda_0=\Delta$ being their conformal weight (we take $H_0=L_0$
for simplicity). Since $\phi,\psi$ are the conformal shadows of states which
are degenerate for all (sufficiently large) values of $N$, we may assume that
 the Jordan cells are preserved by the perturbation, to any order, that is,
\be
(H - \lambda) \Phi = 0, \qquad (H - \lambda) \Psi = \Phi.
\ee
Expanding in power series in $g$, $\lambda=\Delta+g\lambda_1+\ldots$,
$\Phi=\phi+g\phi^{(1)}+\ldots$ and $\Psi=\psi+g\psi^{(1)}+\ldots$, we obtain
at first order the consistency condition $\la \phi|V|\phi \ra = 0$ and
\be
\beta \lambda_1 = \la \phi|V|\psi\ra = \la \psi|V|\phi\ra.
\ee
Since $\psi$ is the only state to have a non-zero scalar product with
$\phi$, the r.h.s. is proportional to $\beta$, and so $\lambda_1$ does
not depend on $\beta$.

The required matrix element can be computed by the procedure used in
\cite{reinicke87} for highest weight representations. To avoid confusions,
we denote hereafter by $(\phi_k,\psi_k)$ the lowest logarithmic pair
in ${\cal R}_k$, with conformal weight $h$. Thus we view the pair $(\phi,\psi)$
as descendants fields at level $r$ of $(\phi_k,\psi_k)$, $\Delta = h + r$.

The main idea is to compare the chiral 3-point function on a strip of width $N$,
\be 
F_{\psi_k\varphi\psi_k} (w_i) = \la \psi_k(w_1) \, L_{-2}^2 (w_2) \, \psi_k(w_3) 
\ra_{\rm strip} \label{3pt} 
\ee
with its expression in the operator formalism. The previous
correlator is first computed on the upper-half plane and then
conformally transformed to the strip. It can be shown that the
logarithmic part of $F_{\psi_k\varphi\psi_k}$ is 
\be
F_{\psi_k\varphi\psi_k} (w_i)\Big|_{\rm log} = \beta_k
\Big\{\log\frac{z_3}{z_1} - 2 \log{\Big(1-\frac{z_3}{
z_1}\Big)}\Big\} F_{\pi\varphi\pi}(z_i), \label{Fpfp} 
\ee 
where $F_{\pi\varphi\pi}$ is the strip correlator (\ref{3pt}) with $\psi_k$
replaced by a primary field $\pi$ of the same weight $h$ as $\psi_k$,
and where $z_i = \exp{({\pi w_i/ N})}$ relates the planar
variables to the strip coordinates. The calculation of $F_{\pi \varphi \pi}$ is 
standard but requires the transformation law of the descendant field $L_{-2}^2(w)$. 
Using the results of \cite{gaberdiel94} on the transformation laws of general 
descendant fields, one finds the following explicit function of $x=z_2/z_1$ and $y=z_3/z_2$,
\bea
&& \hspace{-1.5cm} F_{\pi\varphi\pi}(z_i) = \Big(\frac{\pi}{N}\Big)^{2h + 4} \: \frac{(xy)^h}{(1-xy)^{2h}} \nonumber\\
&& \hspace{-10mm}  \times \: \Big\{h(h+2) \frac{(1-xy)^4}{(1-x)^4(1-y)^4} 
- 3h \frac{(1-xy)^2(1+xy)}{(1-x)^3(1-y)^3} \nonumber\\
&& \hspace{-3mm} - h \frac{c-10}{12} \frac{(1-xy)^2}{(1-x)^2(1-y)^2} + \frac{c}{24} \frac{5c+22}{120} \Big\}.
\eea

In the operator formalism, and for $w=u+iv$, the correlator is 
\be
F_{\psi_k\varphi\psi_k} = \la 0|\hat \psi_k(0,v_1) \, T^{u_1-u_2} \,
\hat \varphi(0,v_2) \, T^{u_2-u_3} \, \hat \psi_k(0,v_3)|0 \ra,
\ee 
with the transfer matrix $T = {\rm e}^{-\pi L_0/N}$. Inserting
a complete set of states of ${\cal R}_k$, the contribution coming
from the pair $(\phi,\psi)$ reads 
\be 
F \sim \la 0|\hat
\psi_k(0,v_1) T^{u_1-u_2} \,{\mathbb P} \,\hat \varphi(0,v_2) \,{\mathbb P}
\, T^{u_2-u_3} \hat \psi_k(0,v_3)|0 \ra, \label{psi} 
\ee 
where 
\be
{\mathbb P} = \frac{|\psi\ra\la\phi|}{\beta} + \frac{|\phi\ra\la\psi|
}{ \beta} - \gamma \frac{|\phi\ra\la\phi| }{ \beta^2} 
\ee 
is a
projector on the subspace spanned by $\phi,\psi$, namely
${\mathbb P}|\psi\ra = |\psi\ra$, ${\mathbb P}|\phi\ra = |\phi\ra$ and
${\mathbb P}|n\ra = 0$ for any other state $|n\ra$. Using the action of
the transfer operator, namely ${\rm e}^{xL_0} |\phi\ra = {\rm
e}^{x\Delta} |\phi\ra$ and ${\rm e}^{xL_0} |\psi\ra = {\rm
e}^{x\Delta} (|\psi\ra + x |\phi\ra)$, we find that (\ref{psi})
has a term quadratic in the $u$'s, 
\bea 
&& \hspace{-1cm}\frac{\pi^2 }{ \beta^2N^2} \, {\rm e}^{- \frac{\pi}{N}\Delta(u_1-u_3)} \,
(u_1-u_2)(u_2-u_3) \nonumber\\
&& \hspace{-5mm} \times \la 0|\hat\psi_k(0,v_1)|\phi\ra \, \la
\phi|\hat\varphi(0,v_2)|\phi\ra \, \la \phi|\hat\psi_k(0,v_3)|0\ra.
\eea
As $u \sim \log z$, this term would require a corresponding $\log^2$
term in the 3-point function $F_{\psi_k\varphi\psi_k}$, which does not exist.
We thus have $\la \phi|\hat\varphi(0,v_2)|\phi\ra = 0$, which fulfills
the consistency condition we found in perturbation theory.

Looking now at the terms linear in the $u$'s, we find
\bea
&&\hspace{-1.4cm} -\frac{\pi}{ \beta^2 N} {\rm e}^{-\frac{\pi }{N}\Delta (u_1-u_3)} \times \nonumber\\
&&\hspace{-1.2cm} \Big\{ (u_1-u_2) \la 0|\hat\psi_k(0,v_1)|\phi\ra \,
\la \phi|\hat\varphi(0,v_2)|\psi\ra \, \la \phi|\hat\psi_k(0,v_3)|0\ra \nonumber\\
&&\hspace{-1.2cm} + (u_2-u_3) \la 0|\hat\psi_k(0,v_1)|\phi\ra \, \la
\psi|\hat\varphi(0,v_2)|\phi\ra \, \la
\phi|\hat\psi_k(0,v_3)|0\ra\Big\}.\nonumber\\
\eea 
The dependence in $u_2$ must vanish, which implies the equality $\la
\phi|\hat\varphi(0,v_2)|\psi\ra = \la
\psi|\hat\varphi(0,v_2)|\phi\ra$, so that a single term remains, 
\bea
&&\hspace{-1.4cm} -\frac{\pi}{\beta^2 N} \, {\rm e}^{-\frac{\pi}{
N} \Delta(u_1-u_3)} \, (u_1-u_3)
\times \nonumber\\
&& \hspace{-5mm} \la 0|\hat\psi_k(0,v_1)|\phi\ra \, \la
\phi|\hat\varphi(0,v_2)|\psi\ra \, \la
\phi|\hat\psi_k(0,v_3)|0\ra. 
\eea 
Since ${\rm e}^{-\frac{\pi}{N}(u_1-u_3)} \sim {z_3/z_1}$, a comparison with (\ref{Fpfp})
implies, provided the pair $(\phi,\psi)$ is not degenerate, that
the product of the three matrix elements in the previous equation
equals the coefficient of $(z_3/z_1)^\Delta = (xy)^\Delta$ in
$\beta^2\beta_k F_{\pi\varphi\pi}$, and does not depend
on $v_2$.

Repeating the same steps for the 2-point function 
\be
F_{\psi_k\psi_k}(w_1,w_3) = \la \psi_k(w_1) \, \psi_k(w_3) \ra_{\rm strip},
\ee
we obtain similarly that its logarithmic part is given by
\be
F_{\psi_k\psi_k} (w_i)\Big|_{\rm log} = \beta_k
\Big\{\log\frac{z_3}{z_1} - 2 \log{\Big(1-\frac{z_3}{
z_1}\Big)}\Big\} F_{\pi\pi}(z_i), 
\ee 
where $F_{\pi\pi}$ is the strip 2-point function for a primary field of weight $h$,
\be
F_{\pi\pi}(z_i) = \Big(\frac{\pi}{N}\Big)^{2h} \: \frac{(z_1z_3)^h}{(z_1 - z_3)^{2h}},
\ee
a function of the single argument $xy$.

Comparing this function with its corresponding expression in the operator formalism, we can conclude that the product $\la 0|\hat\psi_k(0,v_1)|\phi\ra \, \la \phi|\hat\psi_k(0,v_3)|0\ra$
is equal to the coefficient of $(z_3/z_1)^\Delta$ in $\beta\beta_k F_{\pi\pi}$.

Combining the above two results on the triple and double products, we obtain that their ratio yields 
\be
\la \phi|\hat\varphi(0,v_2)|\psi\ra = \la \phi|\hat\varphi(0,0)|\psi\ra = \beta
\frac{F_{\pi\varphi\pi}(x,y) \Big|_{(xy)^\Delta}}{
F_{\pi\pi}(xy)\Big|_{(xy)^\Delta}}. 
\label{matelement}
\ee 
This formula shows
that the first correction $\lambda_1$ to the conformal energy of
$\psi$ can be computed as if $\psi$ was a (non-degenerate)
descendant $|\Delta\ra$ of a primary field $\pi$,
\bea
\lambda_1 \equal \frac{1}{\beta} \la \phi|V|\psi\ra = \frac{1}{\beta} \int_0^N {\rm d}v \: \la \phi | \hat\varphi(0,v) |\psi\ra \nonumber\\
\equal N \frac{F_{\pi\varphi\pi}(x,y) \Big|_{(xy)^\Delta}}{
F_{\pi\pi}(xy)\Big|_{(xy)^\Delta}} = \la \Delta|V|\Delta\ra.
\eea
Being degenerate with $\psi$, the
first correction for $\phi$ is equal to that of $\psi$. Thus all
Jordan couplings between sectors for $N$ even can be neglected for
the calculation of $\lambda_1$, and each sector ${\cal
L}_{N,\ell}$ can be considered separately, as carrying a highest
weight representation ${\cal V}_{1,s}$. Note that the formula 
(\ref{matelement}) requires $h \neq 0$.

At higher orders in perturbation, we have no general argument to show that the same conclusion
holds. On the other hand, we expect that the first order corrections coming from more irrelevant perturbations 
in the tower of the identity show the same independence to the Jordan structure.

So let $|\Delta\ra = |h,r\ra$ be a non-degenerate, level $r$
descendant state in a representation of highest weight $h$.
The normalized\footnote{The normalization in \cite{reinicke87} is 
$(N/2\pi)^4$ because the calculation is performed on a cylinder, in which case 
the planar and cylinder variables are related by $z = \exp{(2\pi w/N)}$.} 
matrix elements $C_{\Delta\varphi \Delta} \equiv (N/\pi)^4
\la \Delta|\hat\varphi(0,0)|\Delta\ra$ for $\varphi = (L_{-2}^2 {\mathbb
I})$ can be computed from (\ref{matelement}) by using the explicit functions 
$F_{\pi \varphi\pi}$ and $F_{\pi\pi}$ above, and reproduce, for $h \neq 0$, the
results first found in \cite{reinicke87}, 
\bea 
&& \hspace{-12mm}C_{\Delta \varphi \Delta} = (h+r)\Big(h + \frac{r(2 h + r)(5 h+1)
}{ (h+1)(2h+1)} - \frac{2+c}{12}\Big) \nonumber\\
&& \hspace{-8mm} +\: \Big(\frac{c}{24}\Big)^2 + \frac{11c}{1440}
+ \frac{r}{30}\Big[r^2(5c-8)-(5c+28)\Big] \delta_{h,0}.
\label{rei} 
\nonumber\\
\eea 
We obtain from this the first correction to the
conformal energy of a state $|\Delta\ra$ as 
\be
g \lambda_1 = \la \Delta|gV|\Delta\ra = \frac{g \pi^4}{ N^3} C_{\Delta\varphi \Delta}.
\ee

We may now apply this to the two lowest levels $E_0^{(\ell)}$ and $E_1^{(\ell)}$
in each sector ${\cal L}_{N,\ell}$ (the only ones to be generically
non-degenerate). For those states, $h=h_{1,s}$ and $r=0$ or
$r=1$, so that we find from the formula (\ref{rei}) with $c=-2$,
\bea
g\lambda_1 \equal \frac{g \pi^4 }{ N^3} \Big(h_{1,s}^2 - \textstyle {\frac{1 }{ 120}}\Big),\qquad r=0,\\
g\lambda_1 \equal \frac{g \pi^4 }{ N^3} \Big(h_{1,s}^2 + 6 h_{1,s} +
\textstyle \frac{119 }{ 120}\Big),\qquad r=1. 
\eea

The corrections obtained earlier in (\ref{ags}), (\ref{a20}) and
(\ref{a21}) from the lattice transfer matrix are clearly
reproduced by setting $g = \lambda_3(\zeta)/6\pi$. We note that
$c=-2$ is the only value for which the perturbation theory
correctly reproduces the lattice finite-size corrections. If we do
not fix $c$, we see from (\ref{a1}) that the weight $h$ is shifted to $h =
h_{1,s}+{(c+2) \over 24}$, and from the formula (\ref{rei}), the first
order $\lambda_1$ for $r=0$ and $r=1$ gets an extra term, inside
the parentheses, equal to ${(c+2) \over 1440}$ and $241(c+2) \over 1440$
respectively.

\section*{Acknowledgments}

We thank Jorgen Rasmussen for useful comments. This work was
supported by Grants NSC 100-2112-M-001-003-MY2, NCTS (North), and
the Belgian Interuniversity Attraction Poles Program P6/02. P.R.
is Senior Research Associate of the Belgian National
 Fund for Scientific Research (FNRS).

\end{document}